\documentclass{IET}

\usepackage{hyperref} 
\usepackage{graphics} 
\usepackage{algorithmic} 
\usepackage{bm}
\usepackage{graphicx}
\usepackage{subfigure}
\usepackage{amsmath}
\usepackage{ulem}
\usepackage{caption}
\usepackage{amsthm,amssymb,amsmath}
\usepackage{lmodern}
\usepackage{mathrsfs}
\DeclareMathAlphabet{\mathpzc}{OT1}{pzc}{m}{it}
\usepackage{mathtools}

\DeclarePairedDelimiter\floor{\lfloor}{\rfloor}
\usepackage[linegoal]{tabu}\tracingtabu=2

\begin{document}

\title{A Binary Wyner-Ziv Code Design Based on Compound LDGM-LDPC Structures}

\author{Mahdi Nangir}
\affil{{\small Faculty of Electrical Engineering, K.N.Toosi University of Technology, Tehran, Iran. E-mails: mahdinangir@ee.kntu.ac.ir-mahmoud@eetd.kntu.ac.ir}}

\author[1]{Mahmoud Ahmadian-Attari}

\author[2,*]{and Reza Asvadi}
\affil{{\small Faculty of Electrical Engineering, Shahid Beheshti University, Tehran, Iran. E-mail: r\_asvadi@sbu.ac.ir}}
\affil[*]{{\small Corresponding Author: r\_asvadi@sbu.ac.ir}}

\abstract{In this paper, a practical coding scheme is designed for the binary Wyner-Ziv (WZ) problem by using nested low-density generator-matrix (LDGM) and low-density parity-check (LDPC) codes. This scheme contains two steps in the encoding procedure. The first step involves applying the binary quantization by employing LDGM codes and the second one is using the syndrome-coding technique by utilizing LDPC codes. The decoding algorithm of the proposed scheme is based on the Sum-Product (SP) algorithm with the help of a side information available at the decoder side. It is theoretically shown that the compound structure has the capability of achieving the WZ bound. The proposed method approaches this bound by utilizing the iterative message-passing algorithms in both encoding and decoding, although theoretical results show that it is asymptotically achievable.}

\maketitle

\section{Introduction}
\label{intro}
Recently, graph-based codes, i.e., codes with a graphical representation, are mostly utilized in channel coding, source coding, and also in joint source-channel coding schemes because of their superior performance in achieving theoretical bounds and low computational complexity. Turbo codes and LDPC codes with iterative decoding algorithms are able to achieve the capacity of most of the known communication channels \cite{1,2}. In addition, LDGM codes, which are the source code dual of the LDPC codes, perform very well in achieving the rate-distortion bound of the binary-symmetric source \cite{3}, and the binary-erasure source \cite{4}.

Lossy source coding with a side information available at the decoder, known as the Wyner-Ziv (WZ) coding, is a fundamental problem in distributed source coding \cite{17}. This problem also arises in practical applications, e.g., wireless sensor networks, and distributed video coding. The WZ problem in the Gaussian domain has been abundantly studied, however, less attention has been paid to this problem in the binary context. Previous studies have used various types of source and channel coding schemes for dealing with the WZ problem. Briefly, if an efficient combination of a good source code with a performance close to the Shannon rate-distortion limit and a good channel code with a performance close to the channel capacity limit are utilized in this problem, then the WZ limit is achievable \cite{8,9}.

To the best of our knowledge, there are few code designs for the binary WZ problem. Nested Convolutional/Turbo codes are one of the most efficient codes employed in the WZ problem, which have been proposed in \cite{9}. This work has achieved within $0.09$ bits away from the binary WZ limit for Turbo code of length $3 \times {10}^5$. Furthermore, polar codes and spatially-coupled LDPC codes have been used for the Quadratic Gaussian WZ problem in \cite{24}, and for the binary WZ problem in \cite{6}, respectively. Sartipi and Fekri have presented a coding scheme for the binary WZ problem based on the parity approach \cite{mina}. They have achieved $0.2$ bits away from the binary WZ limit for the LDPC codes of length about $1000$.

The aim of this paper is designing an efficient code which approaches the binary WZ limit. Our proposed scheme is in the framework of compound LDGM-LDPC codes. LDGM and LDPC codes can be jointly used to form compound codes which belong to the category of nested codes \cite{18, 7}. Nested codes are applicable in most scenarios of coding theory such as noiseless binning and multi-terminal source coding \cite{7}-\cite{12}. 

A crucial component of each WZ coding problem comprises a quantization part. There exist some binary quantization schemes such as Survey-Propagation and Bias-Propagation (BiP) algorithms which efficiently perform close to the Shannon rate-distortion limit \cite{20, 19}. In our proposed method, which is based on the syndrome approach, the BiP algorithm is employed as a binary quantizer. The proposed construction uses optimized degree distribution of LDPC codes over the Binary-Symmetric Channel (BSC). Furthermore, we have designed LDGM codes based on the optimized LDPC codes, whose variable node degrees follow from Poisson distribution.

The main contribution of this paper is designing a compound structure of LDGM-LDPC codes which performs much closer to the binary WZ limit as compared to the previous studies like \cite{9}, \cite{6} and \cite{mina}. It has been shown in \cite{8} that compound LDPC-LDGM structures asymptotically achieve the binary WZ limit. This paper attempts to design a practical coding scheme for the former information theoretical study by employing efficient message-passing algorithms. Performing close to the theoretical limit in our proposed scheme stems from the low Bit Error Rate (BER) operation of LDPC codes and the performance near the rate-distortion limit of the LDGM code designs. We have achieved about $0.0033$ bits away from the binary WZ limit for the compound LDGM-LDPC codes of length about ${10}^5$ when the correlation between the source and side information is modeled by a BSC with the crossover probability $0.25$.

The rest of this paper is organized as follows. In Section 2, the problem definition and preliminaries of our proposed scheme, including the binary WZ problem, the compound LDGM-LDPC construction, and the syndrome-based decoding scheme are introduced. Next in Section 3, encoding and decoding schemes using message-passing algorithms are described. Then, the design of low-density generator and parity-check matrices are presented by using linear algebra approach and combinatorics. Our method of designing the nested low-density codes is also presented in Section 3. In Section 4, the simulation results and discussions about the rate-distortion performance of the proposed code design and its advantages are given. Finally, Section 5 draws the conclusion and future work.

\section{Preliminaries}
In this paper, vectors and matrices are indicated by lowercase and uppercase boldfaced letters, respectively.
Scalars and realization of random variables are represented by lowercase italic letters. 

The codebook of an individual code is represented by its corresponding generator matrix.
In Tanner graph of codes, the variable and check nodes are depicted by circles and squares, respectively. If all of the variable and check nodes have respectively the same degrees $d_v$ and $d_c$, then their associated matrices are called $(d_v,d_c)$-regular. For irregular codes, two polynomials denoted by $\lambda(x)=\sum_{i = 2}^{D_{v}} \lambda_ix^{i-1}$ and $\rho(x)=\sum_{i = 2}^{D_{c}} {\rho_ix^{i-1}}$ are used for determining their degree distributions from an edge perspective, where $\lambda_{i}$ and $\rho_{i}$ are fractions of edges connected to degree $i$ variable and check nodes, respectively \cite{urban}.

The WZ problem is defined in \cite{17}. Let $\bm{s}=(s_1,...,s_n)$ be a sequence of uniform binary input source; then we deal with a binary WZ problem. An encoding function maps $\bm{s}$ to a lossy compressed vector $\bm{v}$ with length $k$ smaller than $n$. Moreover, a decoding function using $\bm{v}$ and side information $\bm{j}$ with length $n$ decodes $\bm{s}$ to $\bm{\hat s}$. If the correlation between $\bm{s}$ and $\bm{j}$ is modeled by a BSC with the crossover probability of $p$ and distortion is evaluated by the Hamming distance measure, then the rate-distortion theoretical bound for the binary WZ problem is as follows:
\begin{equation}
\label{eq4}
{R_{WZ}}(D) = l.c.e.\{ h(D*p) - h(D),(p,0)\},
\end{equation}
for ${0 \le D \le p}$. In (\ref{eq4}), $D*p=D(1-p)+p(1-D)$ is the binary convolution, $h(x)=-x{\log_2}x-(1-x){\log_2}(1-x)$ is the binary entropy function, and $l.c.e.$ stands for the lower convex envelop of the term ${h(D*p) - h(D)}$ and the point $(p,0)$ in the rate-distortion plane \cite{17}.

The total distortion in our design stems from both encoder and decoder. The encoder distortion is basically related to the mapping of $\bm{s}$ to the nearest (in the sense of Hamming distance) codeword $\bm{x}=(x_1,...,x_n)$ from a codebook, with a syndrome which is necessarily zero; this mapping is called binary quantization. In our design, $\bm{x}$ is selected from a compound LDGM-LDPC codebook. At the decoder, distortion stems from inefficiency which will be negligible if an efficient channel error correcting code is applied with a rate that is smaller than the capacity of the correlation channel. Therefore, the total average distortion is:
\begin{align}
\label{eq6}
{D_t} &= \frac{1}{n}\mathbb{E}[\sum\limits_{i = 1}^n {d({s_i},{{{\hat s}_i}})}]= \frac{1}{n}\mathbb{E}[\sum\limits_{i = 1}^n {d({s_i},{x_i})}] *\frac{1}{n}\mathbb{E}[\sum\limits_{i = 1}^n {d({x_i},{{{\hat s}_i}})}] \buildrel \Delta \over = {d_1}*{d_2},
\end{align}
where $d(.,.)$ is the Hamming distortion measure, and $\mathbb{E}[.]$ denotes the expected value.

\subsection{The Compound LDGM-LDPC Construction}

\begin{figure}[h]
\begin{center}
\centering
\includegraphics[width=3.8in,height=2.5in]{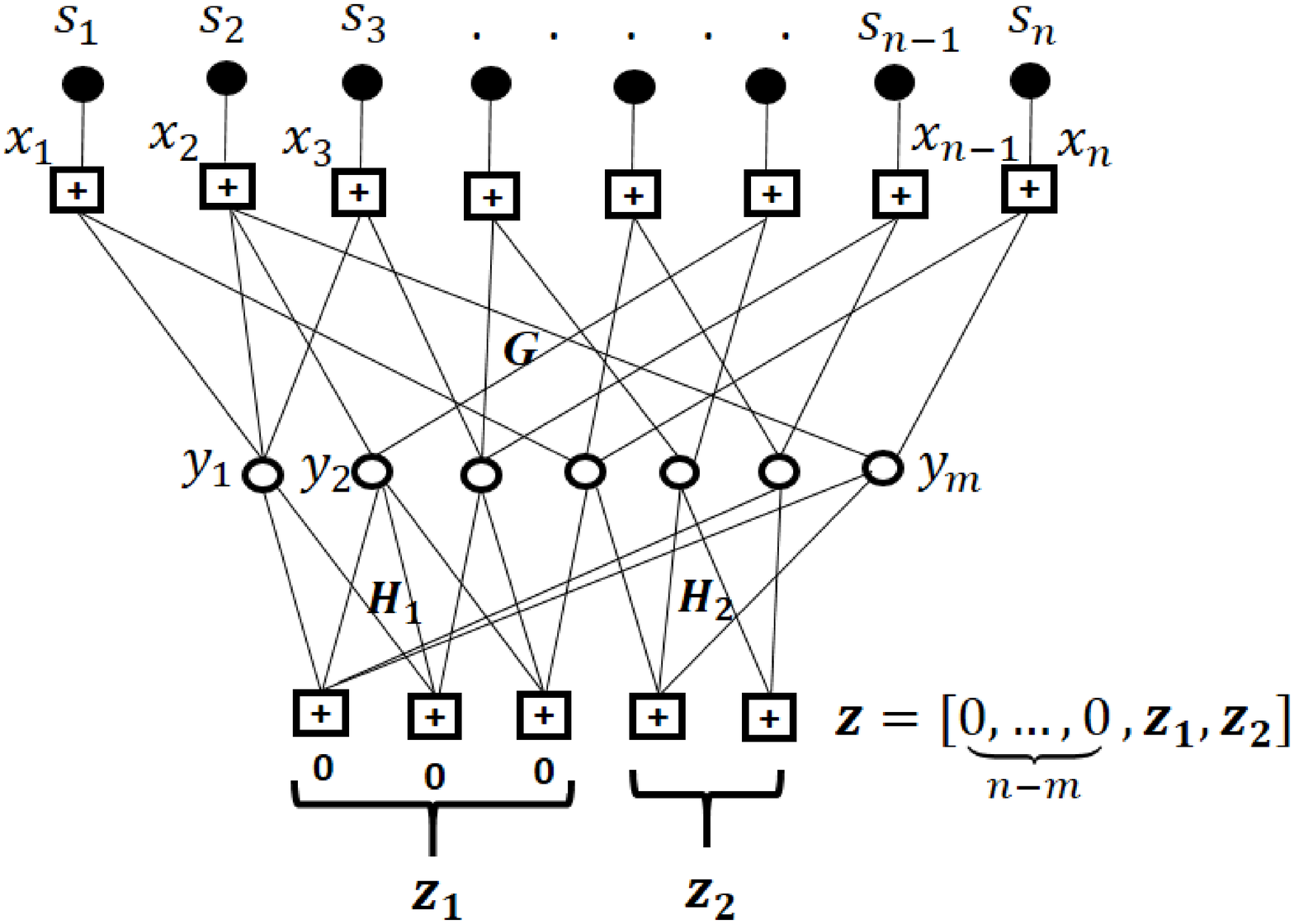}
\end{center}
\vspace{-10pt}
\caption{Tanner graph representation of a compound LDGM-LDPC code construction}
\label{fig3}
\end{figure}

Compound construction consists of an $m \times n$ generator matrix $\bm{G}$ and a parity-check matrix $\bm{\tilde H}\buildrel \Delta \over = [\bm{H_1^T},\bm{H_2^T}]^T$ where $[.]^T$ stands for the matrix transposition. Tanner graph representation of a compound construction is shown in Fig. \ref{fig3}. In this figure, filled circular nodes represent the source nodes. Dimensions of matrices $\bm{H_1}$ and $\bm{H_2}$ are $k_1 \times m$ and $k_2 \times m$, respectively. These matrices result in two nested codes with generator matrices $\bm{\mathcal{G}}$ and $\bm{\mathcal{G}_1}$ via the following relations:

\begin{align}
\label{eqm}
\bm{\mathcal{G}}&=\bm{\tilde G}\times \bm{G},\quad \bm{\mathcal{G}_1}=\bm{G_1}\times \bm{G},
\end{align} 

\hspace{-6.8mm} where $\bm{\tilde G}$ and $\bm{G_1}$ are generator matrices such that $\bm{\tilde G \tilde H^T}=\bm{0}$ and $\bm{G_1 H_1^T}=\bm{0}$. The parity-check matrices of the nested codes $\bm{\mathcal{G}}$ and $\bm{\mathcal{G}_1}$ are denoted by $\bm{\mathcal{H}}$ and $\bm{\mathcal{H}_1}$, respectively. In our proposed scheme, we have considered $\bm{\mathcal{H}}$ to be a parity-check matrix of an LDPC code. Furthermore, some rows of $\bm{\mathcal{H}}$ are chosen to form the submatrix $\bm{\mathcal{H}_1}$ whose generator matrix, $\bm{\mathcal{G}_1}$, has variable node degrees which follow from a Poisson distribution \cite{aref}. 


\begin{figure}[H]
	\begin{center}
		\centering
		\includegraphics[width=5.7in,height=1.1in]{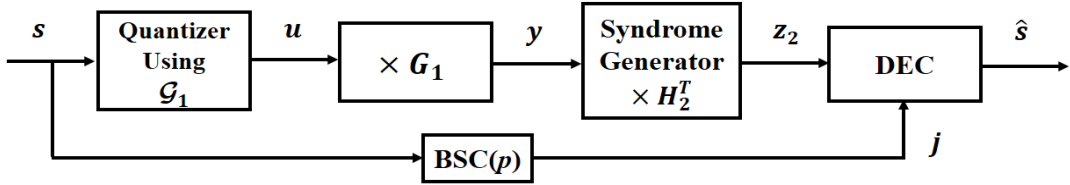}
	\end{center}
	\vspace{-5pt}
	\caption{Block Diagram of the Binary Wyner-Ziv Coding Scheme}
	\label{fig2}
\end{figure}

The block diagram of our coding scheme for the binary WZ problem is illustrated in Fig. \ref{fig2}. It consists of a binary quantization using LDGM codes \cite{16}, which generates $\bm{u}=(u_1,...,u_{m-{k_1}})$ and then $\bm{y}=(y_1,...,y_m)$. Codeword $\bm{x}$ is given by:

\begin{eqnarray}
\label{eq101}
\bm{x}= \bm{u} \times \bm{\mathcal {G}_1}= \bm{u} \times \bm{G_1} \times \bm{G}= \bm{y} \times \bm{G}.
\end{eqnarray}
The syndrome vector of $\bm{y}$ is $\bm{z_1}=\bm{y} \times \bm{H_1^T}=\bm{0}$, and only syndrome $\bm{z_2}=\bm{y} \times \bm{H_2^T}$ is sent to the decoder as a lossy compressed vector of $\bm{s}$ along with the side information $\bm{j}$. In addition to these vectors, the total syndrome $\bm{z}$ is defined as $[\underbrace {0,...,0}_{n-m}, (\bm{z_1}), (\bm{z_2})]$.

At the decoder, a sequence $\bm{\hat s}$ is decoded utilizing the received syndrome $\bm{z_2}$ and the side information $\bm{j}$. This decoding scheme is used for the asymmetric Slepian-Wolf structure \cite{14}. Inspired by the syndrome-decoding idea, if some syndrome bits are set to be zero and only the remaining non-zero bits are sent to the decoder, then a lossy coding scheme is employed for implementation of the binary WZ problem.

\subsection {An Information-Theoretic Viewpoint}

In what follows, the reason why our proposed coding scheme performs close to the rate-distortion limit of the binary WZ problem is briefly explained. Suppose that the source coding rate $R_s$ and the channel coding rate $R_c$ are selected for the binary quantization and the syndrome-coding procedures, respectively, such that the following inequalities are satisfied for arbitrary small positive values of $\varepsilon _s$ and $\varepsilon _c$,
\begin{subequations}
\label{eq1}
\begin{align}
\label{eq1s}
1 - h(d_1) &\le {R_s} < 1 - h(d_1) + {\varepsilon _s}, \\ 
\label{eq1c}
1 - h(d_1*p) - {\varepsilon _c} &< {R_c} \le 1 - h(d_1*p).
\end{align}
\end{subequations}
If a low BER channel decoding algorithm is used in the syndrome-decoding step, then the overall distortion is $D_t = d_1 * d_2\approx d_1$.
In a compound construction, the total rate denoted by $R_t$ equals the difference ${R_s-R_c}$ \cite{8}, and hence:
\begin{align}
\label{eq3aa}
h(D_t*p) - h(D_t)  \le h(d_1*p) - h(d_1) \le R_t< h(d_1&*p)  - h(d_1) + \underbrace {{\varepsilon _s} + {\varepsilon _c}}_\varepsilon  \nonumber \\
&\approx h(D_t*p) - h(D_t) + \varepsilon,
\end{align}
because $d_1 \le D_t$, and $h(x*p) - h(x)$ is a non-increasing function of $x$. Therefore, if ${R_s}$ and ${R_c}$ are chosen according to (\ref{eq1}), then $R_t$ becomes arbitrarily close to $h(D_t*p) - h(D_t)$, because $\varepsilon = {\varepsilon _s} + {\varepsilon _c}$ can be arbitrarily small by a proper design of source and channel codes in the compound structure. Consequently, this leads to achieving the binary WZ bound (\ref{eq4}).

\section {The Proposed Scheme}

\subsection{The Encoding Algorithm}
The compression procedure of $\bm{s}$ consists of two steps:
(Step 1) The source sequence $\bm{s}$ with length $n$ is quantized to a codeword $\bm{{x}} \in \bm{\mathcal {G}_1}$ by using the BiP algorithm. Let $\bm{u}$ denote the information bits after quantization. Next, $\bm{y} =\bm {u} \times \bm{G_1}$ and $\bm{x} = \bm{y} \times \bm{G}$ are calculated.
(Step 2) The syndrome $\bm{z_2} = \bm{y}\bm{H_2^T}$ is obtained, and then it is sent to the decoder. Thus, the overall compression rate in this scheme is ${k_2} \over n$.
Consider the following example for further clarification. For simplicity, short length vectors and small-size matrices are considered.

\textit{Example 1:} Suppose that the source sequence is $\bm{s}=(1,0,0,1,1,0,0,1,0,0)$ with length ${n=10}$. The following matrices in the compound LDGM-LDPC construction are also considered with the size of $m \times n$, $k_1 \times m$, and $k_2 \times m$, respectively, for $\bm{G}$, $\bm{H_1}$, and $\bm{H_2}$ matrices, where $m=8$, $k_1=4$, and $k_2=2$.

\[
\resizebox{.4\textwidth}{!}{\mbox{\ensuremath{\displaystyle
\bm{G}= 
\begin{pmatrix}
1 & 1 & 1 & 1 & 1 & 1 & 1 & 1 & 1 & 1 \\
1 & 1 & 1 & 1 & 1 & 1 & 1 & 1 & 1 & 1 \\
1 & 1 & 1 & 1 & 1 & 1 & 1 & 1 & 1 & 1 \\
1 & 1 & 1 & 1 & 1 & 1 & 1 & 0 & 1 & 1 \\
0 & 0 & 0 & 0 & 0 & 0 & 1 & 0 & 0 & 0 \\
0 & 0 & 0 & 0 & 0 & 0 & 0 & 1 & 0 & 0 \\
1 & 0 & 1 & 0 & 1 & 0 & 0 & 0 & 1 & 0 \\
0 & 1 & 1 & 0 & 0 & 1 & 0 & 0 & 0 & 1
\end{pmatrix}
}}}
,\]

\[
\resizebox{.5\textwidth}{!}{\mbox{\ensuremath{\displaystyle
\bm{\tilde H}= \begin{pmatrix}
\bm{H_1} \\
\bm{H_2}
\end{pmatrix}
=
\begin{pmatrix}
\left\{
\begin{matrix}
1 & 0 & 0 & 0 & 0 & 0 & 1 & 1 \\
0 & 1 & 0 & 0 & 0 & 0 & 1 & 1 \\
0 & 0 & 1 & 0 & 0 & 0 & 1 & 1 \\
0 & 0 & 0 & 1 & 0 & 0 & 1 & 1
\end{matrix} \right\} \bm{H_1} \\
\left\{
\begin{matrix}
0 & 0 & 0 & 0 & 1 & 0 & 1 & 1 \\
0 & 0 & 0 & 0 & 0 & 1 & 1 & 1
\end{matrix} \right\} \bm{H_2}
\end{pmatrix}
}}}
.\]

The generator matrices of LDPC and LDGM codes are calculated according to (\ref{eqm}). These matrices are, respectively, as follows:

\[
\resizebox{.4\textwidth}{!}{\mbox{\ensuremath{\displaystyle
\bm{\mathcal{G}}= 
\begin{pmatrix}
1 & 0 & 1 & 0 & 1 & 0 & 1 & 0 & 1 & 0 \\
0 & 1 & 1 & 0 & 0 & 1 & 1 & 0 & 0 & 1
\end{pmatrix}
}}}
,\]

\[
\resizebox{.4\textwidth}{!}{\mbox{\ensuremath{\displaystyle
\bm{\mathcal{G}_1}= 
\begin{pmatrix}
0 & 0 & 0 & 0 & 0 & 0 & 1 & 0 & 0 & 0 \\
0 & 0 & 0 & 0 & 0 & 0 & 0 & 1 & 0 & 0 \\
1 & 0 & 1 & 0 & 1 & 0 & 0 & 1 & 1 & 0 \\
0 & 1 & 1 & 0 & 0 & 1 & 0 & 1 & 0 & 1
\end{pmatrix}
}}}
.\]
Furthermore, one of the parity-check matrices associated with the LDPC code $\bm{\mathcal{G}}$ is given by:

\[
\resizebox{.4\textwidth}{!}{\mbox{\ensuremath{\displaystyle
\bm{\mathcal{H}}=
\begin{pmatrix}
1 & 0 & 0 & 0 & 1 & 0 & 0 & 0 & 0 & 0 \\
0 & 1 & 0 & 0 & 0 & 1 & 0 & 0 & 0 & 0 \\
0 & 0 & 1 & 0 & 0 & 0 & 1 & 0 & 0 & 0 \\
0 & 0 & 0 & 1 & 0 & 0 & 0 & 1 & 0 & 0 \\
0 & 0 & 0 & 0 & 1 & 0 & 0 & 0 & 1 & 0 \\
0 & 0 & 0 & 0 & 0 & 1 & 0 & 0 & 0 & 1 \\
1 & 0 & 1 & 0 & 0 & 0 & 0 & 1 & 0 & 1 \\
1 & 0 & 0 & 1 & 1 & 0 & 0 & 0 & 0 & 0
\end{pmatrix}
}}}
.\]

Using these matrices, the compressed sequence $\bm{z_2}$ is obtained from the source sequence $\bm{s}$. The LDGM code $\bm{\mathcal{G}_1}$ has $2^{(m-k_1)}=16$ codewords with length $n$. The nearest codeword to $\bm{s}$ is found by using a simple exhaustive search. Therefore, $\bm{s}$ is quantized to the codeword $\bm{x}=(1,0,1,0,1,0,0,1,1,0)$, which is the nearest codeword to $\bm{s}$ regarding the Hamming distance. The message sequence associated with $\bm{x}$ is $\bm{u}=(0,0,1,0)$. By having $\bm{u}$, the lossy compressed sequence $\bm{z_2}$ is calculated,

\begin{align}
\label{eq801}
\bm{y}&=\bm{u} \bm{G_1}=(1,1,1,1,0,0,1,0), \nonumber \\
\bm{z_{2}}&=\bm{y} \bm{H_2^T}=(1,1). \nonumber
\end{align}

Finding the nearest codeword to $\bm{s}$ is done by the BiP algorithm \cite{19}.
Suppose that an LDGM code with the associated Tanner graph is given. Input bits of $\bm{s}$ are located in the source nodes of its Tanner graph. In each round of the algorithm, bias values are calculated for the variable nodes and they are compared with a threshold value of $t$, where $0<t<1$. According to this comparison, the values of some variable nodes are fixed when their absolute bias values become greater than $t$. If the bias value is positive, then the associated variable bit will be fixed to $0$, and otherwise it will be fixed to $1$. If there are no absolute bias values greater than $t$, then only a variable node with the maximum absolute bias value is fixed. This process continues until all the variable nodes are fixed. Finally, $\bm{u}$ is obtained from the variable nodes. In the $l$-th iteration of each round, the message that is sent from a check node $c_a$ to a variable node $v_i$ is:
\begin{equation}
\label{eq8}
\phi_{{c_a} \to {v_i}}^{(l)} = \prod\limits_{{v_j} \in \bar{\mathcal{N}}(c_a)\backslash \{ v_i\} } {\theta_{{v_j} \to {c_a}}^{(l)}},
\end{equation}
where ${\bar{\mathcal{N}}(c_a)}$ is the set of nodes that are connected to the check node $c_a$, including the source node $s_a$, and $\theta_{{v_j} \to {c_a}}^{(l)}$ is the message that has been sent from the variable node $v_j$ to $c_a$. Moreover, the message that is sent from $s_a$ to $c_a$ is:
\begin{equation}
\label{eq9}
\theta_{{s_a} \to {c_a}}^{(l)} = {( - 1)^{{s_a}}}\tanh (\gamma ) \in [ - 1,1],
\end{equation}
where $\gamma $ is a real number that depends on the LDGM code rate.
Similarly, the message that is sent from $v_i$ to $c_a$ in the ($l+1$)-th iteration of each round is:
\begin{equation}
\label{eq10}
\theta_{{v_i} \to {c_a}}^{(l + 1)} = {{\prod\limits_{{c_b} \in \mathcal{N}(v_i)\backslash \{ c_a\} } {(1 + \phi_{{c_b} \to {v_i}}^{(l)})} - \prod\limits_{{c_b} \in \mathcal{N}(v_i)\backslash \{ c_a\} } {(1 - \phi_{{c_b} \to {v_i}}^{(l)})} } \over {\prod\limits_{{c_b} \in \mathcal{N}(v_i)\backslash \{ c_a\} } {(1 + \phi_{{c_b} \to {v_i}}^{(l)})} + \prod\limits_{{c_b} \in \mathcal{N}(v_i)\backslash \{ c_a\} } {(1 - \phi_{{c_b} \to {v_i}}^{(l)})} }},
\end{equation}
where ${\mathcal{N}(v_i)}$ is the set of check nodes that are connected to $v_i$. The initial bias values $\theta_{{v_i} \to {c_a}}^{(0)} $ are set to be $1$.
Finally, the bias values $\theta_{v_i}$ for each variable node are calculated at the end of each round after $\hat l$ iterations by:
\begin{equation}
\label{eq15}
{\theta_{v_i}} = {{\prod\limits_{{c_b} \in \mathcal{N}(v_i)} {(1 + \phi_{{c_b} \to {v_i}}^{(\hat l)})} - \prod\limits_{{c_b} \in \mathcal{N}(v_i)} {(1 - \phi_{{c_b} \to {v_i}}^{(\hat l)})} } \over {\prod\limits_{{c_b} \in \mathcal{N}(v_i)} {(1 + \phi_{{c_b} \to {v_i}}^{(\hat l)})} + \prod\limits_{{c_b} \in \mathcal{N}(v_i)} {(1 - \phi_{{c_b} \to {v_i}}^{(\hat l)})} }}.
\end{equation}

When the girth of the code is 4, we can equip the BiP algorithm by damping operation to reduce the dependency between messages. Equation (\ref{eq10}) will be changed according to $(3.10)$ and $(3.11)$ in \cite{19}.
\subsection{The Decoding Algorithm}
The decoder receives side information $\bm{j}=\bm{s} \oplus \bm{\nu}$ and syndrome $\bm{z_2}$, which $\oplus$ shows binary addition and $\bm{\nu}=(\nu_1,...,\nu_n)$ is a random vector, where $\nu_i$ for $1\leq i\leq n$ are Bernoulli i.i.d. random variables with parameter $p$. Then, the decoder finds the nearest sequence to $\bm{j}$ in the coset corresponding to syndrome $\bm{z}=[0,...,0, (\bm{z_2})]$, using the Sum-Product (SP) algorithm \cite{23}. This algorithm is performed by an LDPC code with the parity-check matrix $\bm{\mathcal {H}}$. The following example illustrates the decoding procedure of Example 1.

\textit{Example 2:} The syndrome $\bm{z_{2}}=(1,1)$ is received, and the syndrome $\bm{z_{1}}=\bm{y}\times \bm{H_1^T}$ equals $(0,0,0,0)$. The systematic form of the parity-check matrix of the LDPC code, which is used for designing the compound code, is as follows:

\[
\resizebox{.4\textwidth}{!}{\mbox{\ensuremath{\displaystyle
\bm{\mathcal{H}}_{\bf sys}= 
\begin{pmatrix}
1 & 0 & 0 & 0 & 0 & 0 & 0 & 0 & 1 & 0 \\
0 & 1 & 0 & 0 & 0 & 0 & 0 & 0 & 0 & 1 \\
0 & 0 & 1 & 0 & 0 & 0 & 0 & 0 & 1 & 1 \\
0 & 0 & 0 & 1 & 0 & 0 & 0 & 0 & 0 & 0 \\
0 & 0 & 0 & 0 & 1 & 0 & 0 & 0 & 1 & 0 \\
0 & 0 & 0 & 0 & 0 & 1 & 0 & 0 & 0 & 1 \\
0 & 0 & 0 & 0 & 0 & 0 & 1 & 0 & 1 & 1 \\
0 & 0 & 0 & 0 & 0 & 0 & 0 & 1 & 0 & 0
\end{pmatrix}
}}}
.\]
Therefore, $\bm{z}$ equals $\bm{x}  \bm{\mathcal{H}_{\bf sys}^T} = (0,0,0,0,0,0,1,1)$, which is the total syndrome that the decoder receives. Besides, suppose that the side information $\bm{j}=(1,0,1,1,1,0,0,1,0,1)$ is available at the decoder.

The task of the decoder is finding the nearest sequence to the side information $\bm{j}$, which has the total syndrome $(0,0,0,0,0,0,1,1)$. There are four sequences with length $10$ which have the same syndrome value $\bm{z}$. They are:
\begin{align}
\label{eq810}
\bm{a}=(0,0,0,0,0,0,1,1,0,0), \nonumber \\
\bm{b}=(0,1,1,0,0,1,0,1,0,1), \nonumber \\
\bm{c}=(1,0,1,0,1,0,0,1,1,0), \nonumber \\
\bm{d}=(1,1,0,0,1,1,1,1,1,1). \nonumber
\end{align}
The nearest sequence to the side information $\bm{j}$ is $\bm{c}$, so $\bm{\hat s}$ is equal to $\bm{c}$. Note that $\bm{\hat s}=\bm{x}$ declares that there is no distortion in the decoding part.

\subsection{Code Design}
In this subsection, design procedure of low-density graph-based codes in the compound construction is described. In this regard, some definitions and lemmas are provided.

\textit{Definition 1:} The \textit{diagonal} elements of an $m \times n$ matrix ($m \le n$) consist of its $(i,i)$-th entries, for $i=1,2,...,m$.

\textit{Lemma 1:} The degree distribution of a binary matrix $\bm{\mathcal{A}}$ remains the same under any column and/or row interchange.


\textit{Definition 2:} An \textit{all-one-diagonal} binary matrix is a matrix in which the main \textit{diagonal} entries are one and the other entries, i.e., the \textit{off-diagonal} entries, are arbitrary bits. Similarly, an \textit{all-zero-diagonal} binary matrix is a matrix with zero-valued main \textit{diagonal} entries.

\textit{Lemma 2:} For any binary full-rank matrix $\bm{\mathcal{A}}$, there is at least one permutation of columns and/or rows, such that the resulting matrix is an \textit{all-one-diagonal} matrix.

\textit{Proof:} See Appendix A.

\textit{Lemma 3:} Let $\bm{\mathcal{I}}$ be an $m \times n$ binary \textit{all-one-diagonal} matrix, and all of the \textit{off-diagonal} entries are zero, i.e, rectangular identity matrix. Let also $\bm{\mathcal{A}}$ be an $m \times n$ \textit{all-one-diagonal}, binary full-rank matrix with degree distribution $(\lambda (x), \rho (x))$. Suppose that $\bm{\mathcal{A}_0}$ obtains from $\bm{\mathcal{A}}$ by inserting zero-valued entries in the main \textit{diagonal}. Then the following $2m \times 2n$ matrix $\bm{\mathcal{H}}$ has the same degree distribution $(\lambda (x), \rho (x))$,

\[
\resizebox{.2\textwidth}{!}{\mbox{\ensuremath{\displaystyle
\bm{\mathcal{H}}= 
\begin{pmatrix}
\bm{\mathcal{I}} & \bm{\mathcal{A}_0} \\
\bm{\mathcal{A}_0} & \bm{\mathcal{I}}
\end{pmatrix}
}}}
.\]

\textit{Proof:} See Appendix B.

%
In our proposed scheme, an ${{(n-m+k_1+k_2)} \over 2} \times {n \over 2}$ full-rank parity-check matrix is selected for a specified degree distribution $(\lambda (x), \rho (x))$. Then, it is transformed into an \textit{all-one-diagonal} matrix $\bm{\mathcal{A}}$ with the same degree distribution by using proper row and/or column permutations. According to Lemma 1 and Lemma 2, there exists at least one matrix with these properties. 

The ${{(n-m+k_1+k_2)}} \times {n}$ parity-check matrix $\bm{\mathcal{H}}$ of an LDPC code in the compound structure is formed with the same degree distribution $(\lambda (x), \rho (x))$ by using Lemma 3. If $\bm{\mathcal{I}}$ is an ${{(n-m+k_1+k_2)} \over 2} \times {n \over 2}$ rectangular identity matrix, then the first $n-m+k_1$ rows of $\bm{\mathcal{H}}$ are chosen, in order to obtain a submatrix $\bm{\mathcal{H}_1}$ as a parity-check matrix of the LDGM code $\bm{\mathcal{G}_1}$. Suppose the remaining $k_2$ rows are placed in another submatrix $\bm{\mathcal{H}_2}$. If $(n-m+k_1) \le {k_2}$, then $\bm{\mathcal{H}_1}$ is an \textit{all-one-diagonal} matrix. Since, there are several generator matrices for a given parity-check matrix $\bm{\mathcal{H}_1}$, obtaining an $(m-k_1) \times n$ sparse generator matrix $\bm{\mathcal{G}_1}$ is desired such that the weight of its rows follows a Poisson distribution function.

Obviously, the parity-check matrix $\bm{\mathcal{H}_1}$ is in the form of $\begin{pmatrix}  \bm{I} & \bm{O} & \bm{B} \end{pmatrix}$, where $\bm{I}$ and $\bm{O}$ are respectively an $(n-m+k_1) \times (n-m+k_1)$ identity matrix, and an $(n-m+k_1) \times (m-k_1-{n\over 2})$ all-zero matrix. In addition, $\bm{B}$ consists of the first $n-m+k_1$ rows of $\bm{\mathcal{A}_0}$ and it is an $(n-m+k_1) \times {n \over 2}$ matrix.

For designing $\bm{\mathcal{G}_1}$, it is necessary and sufficient to find $m-k_1$ linearly independent codewords of it.
Assume $n$-tuple $\bm{c}_j$ is a codeword of $\bm{\mathcal{G}_1}$, for $j=1,2,...,m-k_1$, i.e., $\bm{\mathcal{H}_1} \bm{c}_j^T=\bm{0}^T$. Split the codeword $\bm{c}_j$ into two \textit{message} parts $\bm{m}_{j,1} \in \{0,1\}^{m-k_1-{n\over 2}}$ and $\bm{m}_{j,2} \in \{0,1\}^{{n\over 2}}$, and a \textit{parity} part $\bm{p}_j \in \{0,1\}^{n-m+k_1}$, such that $\bm{c}_j=(\bm{p}_j,\bm{m}_{j,1},\bm{m}_{j,2})$.
To design the rows of $\bm{\mathcal{G}_1}$, $\bm{m}_j \buildrel \Delta \over =(\bm{m}_{j,1},\bm{m}_{j,2})$ is filled with ${m-k_1}$ different vector of information bits, whose Hamming weight is $w_H(\bm{m}_j)=\zeta$, where $w_H(\bm{b})$ denotes the Hamming weight of the vector $\bm{b}$. Then, the \textit{parity} part $\bm{p}_j$ satisfies $\bm{I}\bm{p}_{j}^T=\begin{pmatrix} \bm{O} & \bm{B} \end{pmatrix}\bm{m}_{j}^T=\bm{B} \bm{m}_{j,2}^T$. Therefore, $\bm{p}_j$ equals binary sum of $\zeta$ columns of the submatrix $\begin{pmatrix} \bm{O} & \bm{B} \end{pmatrix}$. Suppose these columns, denoted by $\bm{b}_{q_{j,i}}$ for $j=1,2,...,m-k_1$, $i=1,2,...,\zeta$, and $q_{j,i} \in \{1,2,...,m-k_1\}$ are sorted such that $w_H(\mathbin{\oplus}_{i=1}^\zeta \bm{b}_{q_{j,i}}) \le w_H(\mathbin{\oplus}_{i=1}^\zeta \bm{b}_{q_{j+1,i}})$ for $j=1,2,...,m-k_1-1$, where $\mathbin{\oplus}_{i=1}^\zeta$ shows the binary sum over index $i$.
Let $p(i)={e^{-\lambda}{\lambda}^i \over {i!}}$ be the probability mass function of a Poisson distribution with parameter $\lambda$, for $i=0,1,2,...$ . Also consider $\{a _j\}_{j=1}^{m-k_1}$ is a monotonically increasing sequence of ordered integer numbers, with $a_j \in \{1,2,...,i_{max}\}$ for a fixed integer number $i_{max}$.
In this sequence, the probability of occurrence of any integer number $i \in \{1,2,...,i_{max}\}$ equals $p(i)$, or equivalently, the number of occurrences of $i$ is $n_i=[p(i) \times (m-k_1)]$, where $[x]$ shows the nearest integer number to $x$. Obviously, $\sum_{i=1}^{i_{max}} n_{i} \approx m-k_1$. By these assumptions, the Hamming weight of $\bm{m}_{j,1}$ is set as follows:

\begin{equation}
\label{weigth}
w_H(\bm{m}_{j,1})=\floor*{a_j-(w_H(\mathbin{\oplus}_{i=1}^\zeta \bm{b}_{q_{j,i}})+\zeta)}_{+},
\end{equation}
where $\floor*{x}_{+}$ is $x$ for $x \ge 0$, and otherwise is zero. Then, the positions of $1$'s in $\bm{m}_{j,1}$ are chosen in a way that the resulting vectors be linearly independent.

Note that the Hamming weight of $\bm{m}_{j,1}$ does not affect $\bm{p}_j$. Hence, $\bm{m}_{j,1}$ is filled with some information bits whose weights are chosen according to (\ref{weigth}). By this procedure, the rows of $\bm{\mathcal{G}_1}$ are linearly independent codewords $\bm{c}_j$, whose Hamming weights satisfy the following inequality, for $j=1,2,...,m-k_1$.
\begin{align}
\label{weigths}
w_H(\bm{c}_{j}) &= w_H(\bm{p}_{j})+w_H(\bm{m}_{j,1})+w_H(\bm{m}_{j,2}) \nonumber \\
&= w_H(\mathbin{\oplus}_{i=1}^\zeta \bm{b}_{q_{j,i}})+\floor*{a_j-(w_H(\mathbin{\oplus}_{i=1}^\zeta \bm{b}_{q_{j,i}})+\zeta)}_{+}+\zeta. 
\end{align}
If $a_j \ge w_H(\mathbin{\oplus}_{i=1}^\zeta \bm{b}_{q_{j,i}}) + \zeta$, then $w_H(\bm{c}_{j}) \le a_j$. Otherwise, if $a_j < w_H(\mathbin{\oplus}_{i=1}^\zeta \bm{b}_{q_{j,i}}) + \zeta$, then $w_H(\bm{c}_{j}) \le \zeta (\max\limits_{l}\{w_H(\bm{b}_l)\} + 1)$. In the latter case, it is sufficient to take the integer number $i_{max}$ such that $i_{max} \ge \zeta (\max\limits_{l}\{w_H(\bm{b}_l)\} + 1)$. Therefore, in both cases $w_H(\bm{c}_{j}) \le i_{max}$. Since these codewords are sparse vectors, $\bm{\mathcal{G}_1}$ will be an LDGM code with the Poisson degree distribution for variable nodes.
There are some limitations regarding the block lengths that should be considered in design and implementation, they are mentioned in the following lemmas.

\textit{Lemma 4:} In our code design, the following inequalities are satisfied in the compound structure,

\label{mm}
\begin{align}
\label{eq810a}
n-k_2\le m-k_1, \quad {n\over 2} \le m-k_1.
\end{align}

\textit{Proof:} See Appendix C.

\textit{Lemma 5:} In the compound structure, if ${k_1} + {k_2} \le m \le 2{k_1} + {k_2}$ is satisfied for the given nested codes $\bm{\mathcal {G}_1}$, $\bm{\mathcal {G}}$, and the parity-check matrix $\bm{\tilde H}$, then the existence of at least one binary matrix $\bm{G}$ is guaranteed.

\textit{Proof:} See Appendix D.

The aforementioned inequalities are actually practical limitations in the selection of block lengths. However, there exist two important information theoretic limitations in our design. Firstly, if the compression rate and the resulting distortion of Step 1 are, respectively, considered to be $R_1$ and $d_1$, then ${R_1} > 1 - h({d_1})$ should be satisfied. Secondly, the rate ${R_2}$ should be smaller than the capacity of correlation channel between the quantized sequence $\bm{x}$ and the side information $\bm{j}$ in the syndrome-coding. In other words, if ${R_2} < 1 - h(p*{d_1})$, then the decoding distortion ${d_2}$ or its associated BER in the syndrome-decoding step can be arbitrarily small. In fact, the main result of our practical coding scheme is based on exploiting these limitations by an efficient code design which is stated in the following theorem.

\textit{Theorem 1:} The compound LDGM-LDPC structure is able to achieve the binary WZ limit if it satisfies both of the following features:
\begin{enumerate}
\item An efficient lossy source coding algorithm is used for the binary quantization by which the rate-distortion limit is achievable.
\item A low BER channel decoding algorithm is applied in the syndrome-decoding step.
\end{enumerate}

\textit{Proof:} See Appendix E.

The BiP and the SP algorithms are able to satisfy both of these conditions in the Theorem 1. However, iterative message-passing algorithms are sub-optimal and there is a slight gap between theoretical limits and the rate-distortion performance of these algorithms.
\section{Numerical Results and Discussion}
In this section, some simulation results are presented to demonstrate the performance of the proposed coding scheme at different rates. Optimized degree distributions of irregular LDPC codes over the BSC are used in our simulations. Additionally, the proposed scheme is also implemented by using regular parity-check matrices. The degree distributions of LDPC codes have been presented in Appendix F
\footnote{The degree distributions are obtained from \cite{21}. The matrices are generated according to the degree distributions by using the progressive edge-growth (PEG) algorithm \cite{PEG}.}. Our results are exhibited for two cases of correlation between $\bm{s}$ and $\bm{j}$ where the parameter $p$ equals $0.25$ and $0.05$. Furthermore, an example of the compound codes is provided for which their parameters are adjusted to those used in \cite{6} and \cite{mina}.

\begin{figure}[h]
	\begin{center}
		\includegraphics[width=6.8in,height=3.5in]{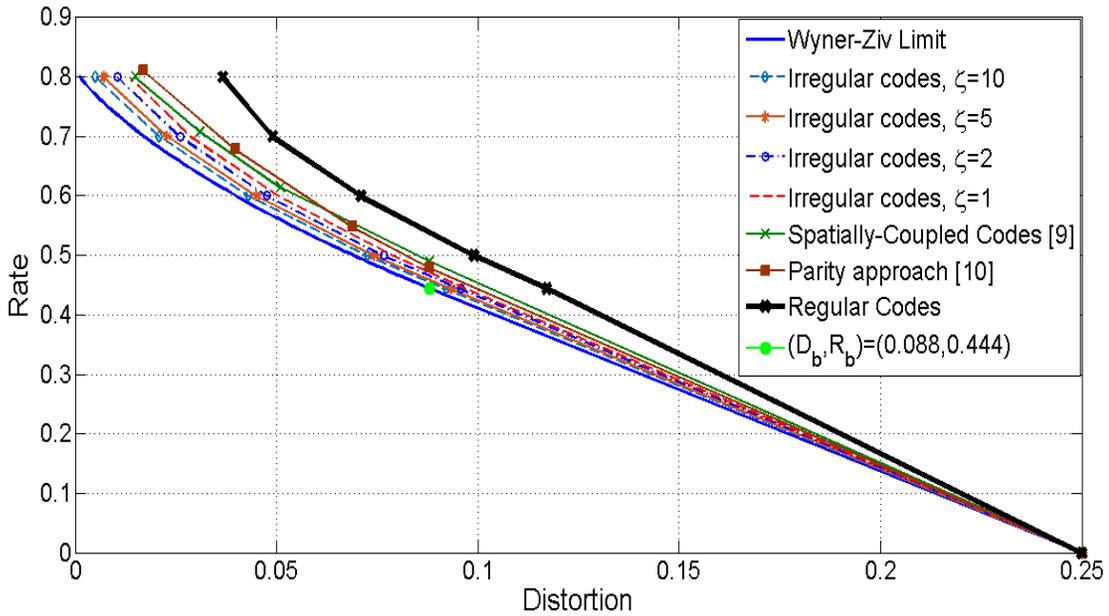}
	\end{center}
	\vspace{-10pt}
	\caption{Rate-Distortion performance of different coding schemes for $p=0.25$ (Example 4)}
	\label{fig9}
\end{figure}

\begin{figure}[h]
\begin{center}
\centering
\includegraphics[width=6.8in,height=3.5in]{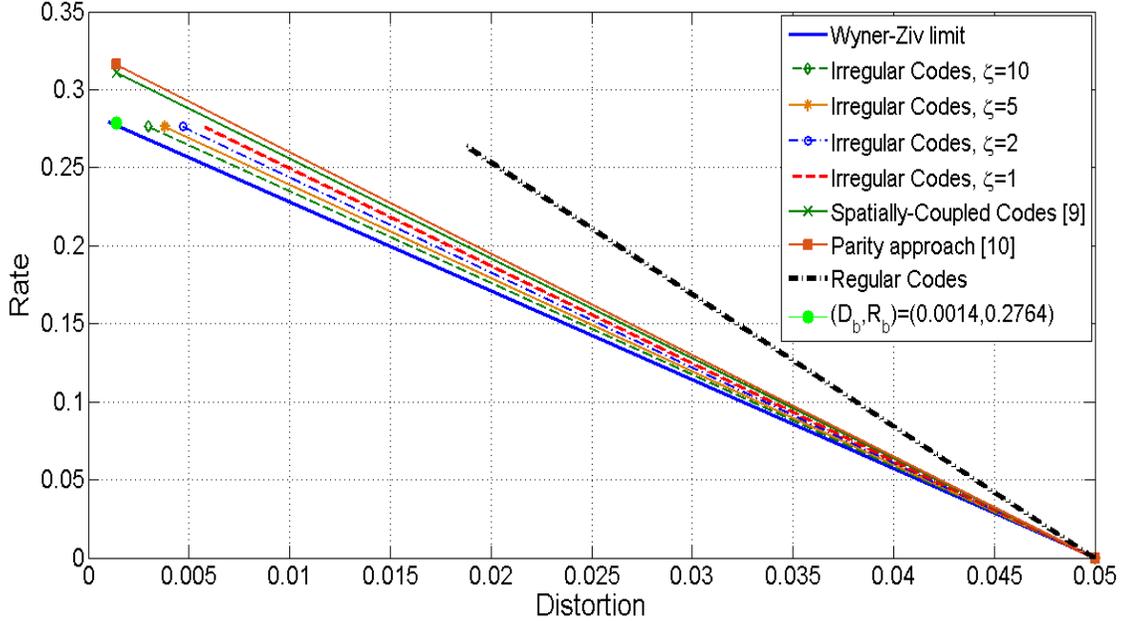}
\end{center}
\vspace{-10pt}
\caption{Rate-Distortion performance of different coding schemes for $p=0.05$ (Example 5)}
\label{fig8}
\end{figure}

The rate-distortion curves of different coding schemes are presented in Figs. \ref{fig9} and \ref{fig8} with the same code length and the same correlation parameter. It is apparent that the rate-distortion performance of the proposed method performs better than other techniques applied in \cite{6} and \cite{mina}. This advantage becomes greater as the parameter $\zeta$ increases as far as the codes remain sparse. For instance, rate-distortion performance of the proposed scheme is depicted for $\zeta=1,2,5$ and $10$.

In implementation of the message-passing algorithms, time sharing between the points of $(p,0)$ and ${({D_b} + \varepsilon_0 ,{R_b})}$ is employed where $({D_b} + \varepsilon_0 ,{R_b})$ is an achieved rate-distortion point, and $(D_b,R_b)$ denotes the boundary point on the binary WZ limit curve. The boundary points are calculated $(0.088,0.444)$ and $(0.0014,0.2764)$ for the correlation parameter values of $0.25$ and $0.05$, respectively. These points separate the WZ limit curve into the high-rate and low-rate regions. The high-rate region is in the form of $h(p*{D_t}) - h({D_t})$ for distortion ${D_t}$, and the low-rate region is a linear curve.

The block size of matrices, the length of codes, code rates, and distortion values are presented in Tables \ref{t2} and \ref{t4} for irregular codes with $\zeta=10$ and in Tables \ref{t2reg} and \ref{t4reg} for regular codes. In these tables, ${d_1}$ and ${R_1}$ indicate distortion and rate of the binary quantization step, respectively. Similarly, ${d_2}$ and ${R_2}$ denote distortion and rate of LDPC codes used in the syndrome-decoding step, respectively. We also use ${D_t}$ and $R_t$ to denote total distortion and rate values, respectively. For more intuition, let $D_{wz}$ be the distortion value of the binary WZ limit at the rate $R_t$. The rates $R_1$, $R_2$, and $R_t$ are calculated as follows:

\label{rates}
\begin{align}
R_1 &= {{m - {k_1}} \over n}, \quad R_2 = {{m - {k_1} - {k_2}} \over n}, \quad R_t = {{k_2} \over n} = {R_1} - {R_2}. 
\end{align}
The distortion values $d_1$, $d_2$, and $D_t$ are obtained from the calculation of the average Hamming distance between the source and the decoded sequences according to (\ref{eq6}). In fact, two types of gaps can be defined to compare the results from rate and distortion point of view. The difference between an achieved distortion (rate) and its associated theoretical limit is called the gap of distortion (gap of rate). The amount of $D_t-D_{wz}$ determines the resulting distortion gap value for a given $R_t$ that is used in the tables. The gap of rate is approximately proportional to the gap of distortion with the ratio $R_b/(p-D_b)$ due to the geometric similarity.


The parameters of $\lambda$ and $i_{max}$ employed in designing the irregular codes are mentioned in Tables \ref{t2} and \ref{t4}.
Each point in our simulations has applied $100$ randomly generated source sequences $\bm{s}$ with uniform distribution. The value of maximum iteration $\hat{l}$ is set to be $25$ in each round of the BiP algorithm. Moreover, we have set $t =0.8$ and $\gamma \approx 2{R_{1}}=2{{m - {k_1}} \over n}$ in our simulations. In addition, the maximum number of iterations in the SP algorithm is set to be $100$.

\textit{Example 4:} Suppose the correlation parameter $p$ equals $0.25$. The block lengths, rates, and distortion values of the compound scheme are presented for irregular and regular codes, respectively, in Table \ref{t2} and Table \ref{t2reg}. The rate-distortion performance of our proposed scheme is also illustrated in Fig. \ref{fig9} for this correlation parameter. For instance, in the case of $R_t$ equals $0.6$, the achieved distortion gap for the proposed scheme is $0.0033$ bits away from the WZ limit. However, the gap values at the same rate are $0.079$ and $0.066$ bits per channel use for coding schemes \cite{mina} and \cite{6}, respectively. 
The equivalent gaps in the sense of distortion are, respectively, $0.0288$ and $0.024$ bits. It is apparent that the distortion gap values for the coding methods in \cite{mina} and \cite{6} are about eight times more than that of the proposed scheme due to inherent defects of their methods.

The coding scheme of \cite{6} has imperfections which lead to a considerable amount of gap, such as: only regular LDGM and LDPC codes are used in that structure
which might be substituted by irregular codes. Moreover, there is a possibility of failure in the algorithm of \cite{6} that causes repetition of the encoding process. Hence, the performance of the scheme is not suitable for short length codes as a result of the attempt at encoding source sequences by such a repetitive process.

In the proposed scheme of \cite{mina} that is based on sending parity bits, using a systematic channel encoder is essential. For this reason, a Gaussian elimination should be applied before encoding that increases the complexity of encoding. The designed LDPC codes in \cite{mina} are based on the MacKay codes which are not optimal codes. Alternatively, degree distribution of the codes can be replaced with optimized codes based on sending either parity bits or some parts of it.

\begin {table}[h]
\centering
\caption{Simulation Results for $p=0.25$ and $\zeta=10$- Irregular Codes (Example 4)}
\label{t2}
\vspace{-15pt}
\begin{center}
	\scalebox{.75}{
		\begin{tabular} {| c | c | c | c | c | c | c | c | c | c | c | c | c |}
			\hline\noalign{\smallskip}
			Code &  $n$ &  $m$ &  ${k_1}$ & $m-{k_1}$ &  ${k_2}$ & $({d_1},{R_1})$ & $({d_2},{R_2})$ & $R_t$ & ${D_t}$ & ${D_{wz}}$ & $\lambda$ & $i_{max}$\\
			\noalign{\smallskip}\hline\noalign{\smallskip}
			1& $100000$ & $76800$ & $20000$ & $56800$ & $44400$& $(0.0892,0.568)$ & $(0.0037,0.124)$& $0.444$& $0.0922$& $0.088$ & $873.2$ & $2000$ \\
			2& $100000$ & $90000$ & $26300$ & $63700$ & $50000$& $(0.0695,0.637)$ & $(0.0036,0.137)$& $0.5$& $0.0726$& $0.0688$ & $855.5$ & $2000$ \\
			3& $100000$ & $95700$ & $20000$ & $75700$ & $60000$& $(0.0403,0.757)$ & $(0.003,0.157)$& $0.6$& $0.0431$& $0.0398$ & $714.95$ & $1600$\\
			4& $100000$ & $107200$ & $20000$ & $87200$ & $70000$& $(0.0181,0.872)$ & $(0.003,0.172)$& $0.7$& $0.021$& $0.017$ & $694.05$ & $1600$\\
			5& $100000$ & $120000$ & $21800$ & $98200$ & $80000$& $(0.0021,0.982)$ & $(0.003,0.182)$& $0.8$& $0.0051$& $0.0011$ & $707.25$ & $1600$\\
			\noalign{\smallskip}\hline
	\end{tabular}}
\end{center}
\end{table}

\begin {table}[h]
\caption{Simulation Results for $p=0.25$- Regular Codes (Example 4)}
\label{t2reg}
\vspace{-5pt}
\begin{center}
\scalebox{.8}{
	\begin{tabular} {| c | c | c | c | c | c | c | c | c | c | c |}
		\hline\noalign{\smallskip}
		Code &  $n$ &  $m$ &  ${k_1}$ & $m-{k_1}$ &  ${k_2}$ & $({d_1},{R_1})$ & $({d_2},{R_2})$ & $R_t$ & ${D_t}$ & ${D_{wz}}$\\
		\noalign{\smallskip}\hline\noalign{\smallskip}
		6& $100000$ & $70000$ & $13600$ & $56400$ & $44400$& $(0.097,0.564)$ & $(0.025,0.12)$& $0.444$& $0.1172$& $0.088$ \\
		7& $100000$ & $90000$ & $28000$ & $62000$ & $50000$& $(0.0793,0.62)$ & $(0.0235,0.12)$& $0.5$& $0.0991$& $0.0688$ \\
		8& $100000$ & $95000$ & $20000$ & $75000$ & $60000$& $(0.049,0.75)$ & $(0.0244,0.15)$& $0.6$& $0.071$& $0.0398$ \\
		9& $100000$ & $105000$ & $20000$ & $85000$ & $70000$& $(0.0283,0.85)$ & $(0.0221,0.15)$& $0.7$& $0.0491$& $0.017$ \\
		10& $100000$ & $120000$ & $25000$ & $95000$ & $80000$& $(0.0133,0.95)$ & $(0.024,0.15)$& $0.8$& $0.0367$& $0.0011$ \\
		\noalign{\smallskip}\hline
\end{tabular}}
\end{center}
\end{table}

\textit{Example 5:} In this example, the correlation parameter $p$ is assumed to be $0.05$. The block lengths, rates, and distortion values of the compound scheme are presented for irregular codes with $\zeta=10$ and for regular codes, respectively, in Table \ref{t4} and Table \ref{t4reg}. The rate-distortion performance of our proposed scheme is also depicted  in Fig. \ref{fig8} for the same correlation parameter. In the boundary point, $R_b=0.2764$, we achieve $0.0016$ bits away from the WZ limit, however the gap values of rate at the same point are $0.0396$ and $0.0347$ bits per channel use for the methods in \cite{mina} and \cite{6}, respectively. The equivalent gaps of distortion are, respectively, $0.007$ and $0.0061$ bits for the methods in \cite{mina} and \cite{6} which are about four times more than that of the proposed method.
\begin {table}[h]
\caption{Simulation Results for $p=0.05$ and $\zeta=10$- Irregular Codes (Example 5)}
\label{t4}
\vspace{-15pt}
\begin{center}
	\scalebox{.72}{
		\begin{tabular} {| c | c | c | c | c | c | c | c | c | c | c | c | c |}
			\hline\noalign{\smallskip}
			Code &  $n$ &  $m$ &  ${k_1}$ & $m-{k_1}$ &  ${k_2}$ & $({d_1},{R_1})$ & $({d_2},{R_2})$ & $R_t$ & ${D_t}$ & ${D_{wz}}$ & $\lambda$ & $i_{max}$\\
			\noalign{\smallskip}\hline\noalign{\smallskip}
			11& $100000$ & $180000$ & $81760$ & $98240$ & $27640$& $(0.0017,0.9824)$ & $(0.0013,0.706)$& $0.2764$& $0.003$& $0.0014$ & $442.7$ & $1000$ \\
			\noalign{\smallskip}\hline
	\end{tabular}}
\end{center}
\end{table}
\begin {table}[H]
\caption{Simulation Results for $p=0.05$- Regular Codes (Example 5)}
\label{t4reg}
\vspace{-5pt}
\begin{center}
	\scalebox{.8}{
		\begin{tabular} {| c | c | c | c | c | c | c | c | c | c | c |}
			\hline\noalign{\smallskip}
			Code &  $n$ &  $m$ &  ${k_1}$ & $m-{k_1}$ &  ${k_2}$ & $({d_1},{R_1})$ & $({d_2},{R_2})$ & $R_t$ & ${D_t}$ & ${D_{wz}}$\\
			\noalign{\smallskip}\hline\noalign{\smallskip}
			12& $100000$ & $180000$ & $82360$ & $97640$ & $27640$& $(0.0035,0.9764)$ & $(0.0154,0.7)$& $0.2764$& $0.0188$& $0.0014$ \\
			\noalign{\smallskip}\hline
	\end{tabular}}
\end{center}
\end{table}

It is apparent from Figs. \ref{fig9} and \ref{fig8} that the rate-distortion performance of our scheme is very close to the binary WZ limit. It is possible that we can achieve a closer rate-distortion performance to the WZ limit for small rates by using time sharing. Nevertheless, the gap value can be further reduced by increasing the block length and the parameter $\zeta$.

\textit{Example 6:} This example presents simulation results of applying the proposed scheme by employing the same parameters used in \cite{mina} and \cite{6}. 
Thus, the code lengths and the correlation parameters, i.e., $p=0.27,0.134$, are chosen according to them. The utilized parameters and the results are presented for the irregular codes with $\zeta=10$ in the first and second row of Table \ref{t5} associated to \cite{mina} and \cite{6}, respectively. The achieved distortion gap by the proposed algorithm are also depicted at the last column of Table \ref{t5} for more comparison. 

In \cite{mina}, a gap of rate $0.155$ bits per channel use away from the WZ limit has been achieved when the code length equals $3000$. This is equivalent to $0.0581$ bits of distortion gap. Instead, the distortion gap of the proposed method is $0.0387$ bits away from the binary WZ limit, where it is about one and a half times less than the one in \cite{mina}. Similarly, in \cite{6}, the achieved gap of rate is $0.0222$ bits per channel use for the code length $140000$. In our scheme, the gap of distortion is $0.0037$ bits for the same parameters. It is equivalent to $0.0151$ bits per channel use of gap in rate. By using time sharing in the linear part of the curve, the gap of rate decreases to $0.0029$ bits per channel use, where it is about eight times less than the one in \cite{6}.
          
\begin{table}[h]          
\caption{Parameters and Results of our scheme in a reverse comparison- $\zeta=10$ (Example 6)}
\label{t5}
\vspace{-15pt}
\begin{center}
	\scalebox{.68}{
		\begin{tabular} {| c | c | c | c | c | c | c | c | c | c | c | c | c | c |}
			\hline\noalign{\smallskip}
			Code & $n$ &  $m$ &  ${k_1}$ & $m-{k_1}$ &  ${k_2}$ & $({d_1},{R_1})$ & $({d_2},{R_2})$ & $R_t$ & ${D_t}$ & ${D_{wz}}$ & $\lambda$ & $i_{max}$ & Gap\\
			\noalign{\smallskip}\hline\noalign{\smallskip}
			$13$ & $3000$ & $2000$ & $548$ & $1452$ & $1212$& $(0.138,0.484)$ & $(0.0148,0.08)$& $0.404$& $0.1487$& $0.11$ & $128.77$ & $300$ & $0.0387$ \\
			\noalign{\smallskip}\hline\noalign{\smallskip}
			$14$ & $140000$ & $180000$ & $57400$ & $122600$ & $67300$& $(0.0173,0.8757)$ & $(0.0025,0.395)$& $0.4807$& $0.0197$& $0.016$ & $836.92$ & $2000$ & $0.0037$ \\
			\noalign{\smallskip}\hline
	\end{tabular}}
\end{center}
\end{table}

It is noteworthy that there exists at least one binary matrix $\bm{G}$ in the compound LDGM-LDPC structure, because the condition  $k_1+k_2<m<2k_1+k_2$ is satisfied for all codes of the Tables according to Lemma 5.

\section{CONCLUSION}

In this paper, we proposed an efficient coding scheme for achieving the binary WZ limit employing a compound LDGM-LDPC code construction, which fulfills a nested coding structure. We utilized optimized degree distribution of LDPC codes and the associated LDGM codes nested with the optimized codes. Furthermore, variable node degrees of LDGM codes are designed to be a Poisson distribution.
In our scheme, if efficient source and channel codes are used with the capability of achieving the rate-distortion and the capacity limits, the performance of the compound code gets closer to the binary WZ limit with any arbitrary precision. We applied the BiP algorithm for the binary quantization using LDGM codes and the SP algorithm for the syndrome-decoding using LDPC codes. By employing these algorithms, our simulation results confirmed that the rate-distortion performance of the proposed scheme is very close to the binary WZ theoretical limit.

Future study may extend the proposed coding scheme and the iterative message-passing algorithms for multi-terminal source coding scenarios. Designing multi-terminal quantization and joint decoding algorithms is considered in our future research study. Moreover, using improved message-passing algorithms for reaching smaller gap from the theoretical limit remains a future research topic.


\section{Appendices}

\subsection{Appendix A: Proof of Lemma 2}

We prove the proposition of this lemma by using the mathematical induction on $m$. Suppose that the entries of matrix $\bm{\mathcal{A}}=\begin{pmatrix} {\mathcal{A}_{j,i}} \end{pmatrix}$ are denoted by $ \mathcal{A}_{j,i}$, for $1 \le j \le m$ and $1 \le i \le n$. This element is located on the $j$-th row and the $i$-th column.


{\bf Basis for the induction:} For $m=1$, $\bm{\mathcal{A}}$ becomes an $1 \times n$ matrix. Since $\bm{\mathcal{A}}$ is full-rank, so it has at least one entry that is $1$. Clearly, it is moved to the first column of $\bm{\mathcal{A}}$ to result in an \textit{all-one-diagonal} matrix.

{\bf Induction hypothesis:} Suppose that for any binary full-rank matrix $\bm{\mathcal{A}}$ with $m-1$ rows, there is at least one permutation of columns and/or rows, such that the resulting matrix is an \textit{all-one-diagonal} matrix, ($m>1$).

{\bf Induction step:} Now, we want to show that the lemma is true for any $m \times n$ binary full-rank matrix $\bm{\mathcal{A}}=(\bm{a}_1,\bm{a}_2,...,\bm{a}_n)$, where $\bm{a}_i$ is the $i$-th column of $\bm{\mathcal{A}}$. Consider a column $\bm{a}_i$, $i \in \{1,2,...,n\}$, whose Hamming weight is $w$, i.e., there are $w$ non-zero entries in this column. Suppose that these $w$ non-zero entries are placed in the rows $\{\bm{r}_{\alpha_1},\bm{r}_{\alpha_2},...,\bm{r}_{\alpha_w}\}$, and the remaining rows of column $\bm{a}_i$ are in the set $\{\bm{r}_{1},\bm{r}_{2},...,\bm{r}_{m}\} \backslash \{\bm{r}_{\alpha_1},\bm{r}_{\alpha_2},...,\bm{r}_{\alpha_w}\}$, where $\{\bm{r}_{1},\bm{r}_{2},...,\bm{r}_{m}\}$ is the set of rows of $\bm{\mathcal{A}}$, and $\backslash$ denotes the set-reduction.

If $w=1$, we remove the row $\bm{r}_{\alpha_1}$ and the column $\bm{a}_i$ from the matrix $\bm{\mathcal{A}}$, the resulting $(m-1) \times (n-1)$ matrix, called $\bm{\mathcal{B}}$, will be full-rank, because all of the elements of the column $\bm{a}_i$ except for the row $\bm{r}_{\alpha_1}$ are zero. According to the induction hypothesis, we can transform $\bm{\mathcal{B}}$ to an \textit{all-one-diagonal} matrix, by a proper permutation of columns and/or rows. Now, it is sufficient to locate $\bm{r}_{\alpha_1}$ to the $m$-th row, and $\bm{a}_i$ to the $m$-th column. Let $\bm{r}_{\alpha_1} {\to} \bm{r}_{m}$ and $\bm{a}_i {\downarrow} \bm{a}_m$ show locating the row $\bm{r}_{\alpha_1}$ in the row $m$, and locating the column $\bm{a}_i$ in the column $m$, respectively.

If $w>1$, we claim that there is at least one row in $\{\bm{r}_{\alpha_1},\bm{r}_{\alpha_2},...,\bm{r}_{\alpha_w}\}$, such that after removing it and the column $\bm{a}_i$, the resulting $(m-1) \times (n-1)$ matrix is full-rank. Otherwise, for any row $\bm{r}_{\alpha} \in \{\bm{r}_{\alpha_1},\bm{r}_{\alpha_2},...,\bm{r}_{\alpha_w}\}$, after removing $\bm{r}_{\alpha}$ and the column $\bm{a}_i$, the remaining matrix, called $\bm{\mathcal{B}_{\alpha}}$, will not be a full-rank matrix. Thus, there is a subset of rows of $\bm{\mathcal{B}_{\alpha}}$, call this subset $\mathcal{S}_{\alpha}$, whose sum of elements is an all-zero vector. Obviously, $\bm{r}_{\alpha} \notin \mathcal{S}_{\alpha}$ for all $\alpha \in \{\alpha_1,\alpha_2,...,\alpha_w\}$, and $\mathcal{S}_{\alpha} \subseteq \{\bm{r}_{1},\bm{r}_{2},...,\bm{r}_{m}\} \backslash \{\bm{r}_{\alpha}\}$. For any $\alpha \in \{\alpha_1,\alpha_2,...,\alpha_w\}$, $\mathcal{S}_{\alpha}$ contains an odd number of rows in $\{\bm{r}_{\alpha_1},\bm{r}_{\alpha_2},...,\bm{r}_{\alpha_w}\}$; unless otherwise, if $\mathcal{S}_{\alpha}$ contains an even number of rows in $\{\bm{r}_{\alpha_1},\bm{r}_{\alpha_2},...,\bm{r}_{\alpha_w}\}$, then the sum of its elements will be all-zero, including the element in the column $\bm{a}_i$. This contradicts with the full-rank assumption of $\bm{\mathcal{A}}$. Hence, $\mathcal{S}_{\alpha_1}$ contains an odd number of rows in $\{\bm{r}_{\alpha_1},\bm{r}_{\alpha_2},...,\bm{r}_{\alpha_w}\}$; e.g., suppose these rows are $\{\bm{r}_{\alpha_{\ell}},...\}$, where $\ell \ne 1$. Similarly, $\mathcal{S}_{\alpha_{\ell}}$ contains an odd number of rows in $\{\bm{r}_{\alpha_1},\bm{r}_{\alpha_2},...,\bm{r}_{\alpha_w}\}$, and $\bm{r}_{\alpha_{\ell}} \notin \mathcal{S}_{\alpha_{\ell}}$. Now, the sum of elements of $\mathcal{S}_{\alpha_{1}}$ and $\mathcal{S}_{\alpha_{\ell}}$ leads to an all-zero vector, including the element in the column $\bm{a}_i$. This also contradicts with the full-rank assumption of $\bm{\mathcal{A}}$. Clearly, $\mathcal{S}_{\alpha_{1}} \neq \mathcal{S}_{\alpha_{\ell}}$, because $\bm{r}_{\alpha_{\ell}} \in \mathcal{S}_{\alpha_{1}}$, but $\bm{r}_{\alpha_{\ell}} \notin \mathcal{S}_{\alpha_{\ell}}$.

Therefore, there exists at least one element in the column $\bm{a}_i$, that is $1$ (consider it is in the row $\bm{r}_j$, i.e., $\mathcal{A}_{j,i}=1$); such that after removing the row $\bm{r}_j$ and the column $\bm{a}_i$, the resulting $(m-1) \times (n-1)$ matrix is full-rank. Now, it is sufficient to apply the induction hypothesis and locate $\bm{r}_j$ to the $m$-th row, and $\bm{a}_i$ to the $m$-th column, i.e., $\bm{r}_j \to \bm{r}_m$ and $\bm{a}_i \downarrow \bm{a}_m$. As a result, an \textit{all-one-diagonal} matrix is built with the dimension $m \times n$.

\subsection{Appendix B: Proof of Lemma 3}

The dimension of matrix $\bm{\mathcal{H}}$ is twice the dimension of matrix $\bm{\mathcal{A}}$. Also, the number of columns or rows with a specific weight in the matrix $\bm{\mathcal{H}}$ are doubled in comparison with those of the matrix $\bm{\mathcal{A}}$. Therefore, the degree distribution remains unchanged.

\subsection{Appendix C: Proof of Lemma 4}

The first inequality (\ref{eq810a}) is true, because $n-m+k_1\le k_2$, and the second inequality is true due to the dimension of matrix $\bm{O}$.

\subsection{Appendix D: Proof of Lemma 5}

The inequality ${k_1} + {k_2} \le m $ is trivial and it is always satisfied due to the dimension of matrix $\bm{\tilde H}$. We demonstrate that if $m \le 2{k_1} + {k_2}$, then there exists at least one choice for $\bm{G}$ with $mn$ unknown elements. The total number of known elements is $n(m-{k_1})+n(m-{k_1}-{k_2})$, which are obtained from (\ref{eqm}). These equations are linearly independent, and they are consistent with each other due to the nesting property of the compound codes. Note that in equations (\ref{eqm}), $\bm{G_1}$ and $\bm{\tilde G}$ are known, because $\bm{\tilde H}$ is given. Therefore, if $m \le 2{k_1} + {k_2}$ then $n(m-{k_1})+n(m-{k_1}-{k_2}) \le mn$, and hence it implies that there exists at least one choice for $\bm{G}$ such that (\ref{eqm}) is satisfied for the given $\bm{\mathcal {G}_1}$, $\bm{\mathcal {G}}$, and $\bm{\tilde H}$. Therefore, the proof is completed.

\subsection{Appendix E: Proof of Theorem 1}

Condition 1 results in (\ref{eq1s}) for a small value of $\varepsilon_s$. Similarly, condition 2 means that (\ref{eq1c}) yields $d_2 \approx 0$ for a small value of $\varepsilon_c$. Therefore, the total rate $R_t=R_s-R_c$ in the compound LDGM-LDPC structure will be close to the binary WZ theoretical bound as in (\ref{eq3aa}).

\subsection{Appendix F: Degree Distribution of LDPC Codes}

The following degree distributions of irregular codes are used in Tables \ref{t2}, \ref{t4}, and \ref{t5} for generating parity-check matrices of LDPC codes. These degree distributions are from the edge perspective, and they are obtained from the density evolution optimization \cite{urban}.

------------------------------------------------------------------------------------------------------------------

Code 1:   $1-R_2=0.876$

$\lambda (x) = 0.3424{x} + 0.165{x}^2 + 0.1203{x}^4
+0.0191{x}^5 + 0.012{x}^6 + 0.1416{x}^{10}
+0.0211{x}^{25} + $

$0.0202{x}^{26} + 0.0185{x}^{34}
+0.0428{x}^{36} + 0.0133{x}^{38} + 0.0021{x}^{39}
+0.0104{x}^{40} + 0.0704{x}^{99}$

$\rho (x) = 0.8{x^3}+0.2{x^4}$

------------------------------------------------------------------------------------------------------------------

Code 2:  $1-R_2=0.863$

$\lambda (x) = 0.3424{x} + 0.165{x}^2 + 0.1203{x}^4
+0.0191{x}^5 + 0.012{x}^6 + 0.1416{x}^{10}
+0.0211{x}^{25} + $

$0.0202{x}^{26} + 0.0185{x}^{34}
+0.0428{x}^{36} + 0.0133{x}^{38} + 0.0021{x}^{39}
+0.0104{x}^{40} + 0.0704{x}^{99}$

$\rho (x) = 0.8{x^3}+0.2{x^4}$

------------------------------------------------------------------------------------------------------------------

Code 3:  $1-R_2=0.843$

$\lambda (x) = 0.3151x + 0.1902{x}^2 + 0.045{x}^4
+0.1705{x}^6 + 0.1405{x}^{17} + 0.0081{x}^{37}
+0.044{x}^{41} + $

$0.0863{x}^{66}$

$\rho (x) = 0.5{x^3}+0.5{x^4}$

------------------------------------------------------------------------------------------------------------------

Code 4:   $1-R_2=0.828$

$\lambda (x) = 0.3038x + 0.1731x^2 + 0.0671x^4
+0.0123x^5 + 0.1341x^6 + 0.0314x^{12}
+0.0108x^{14} + $

$ 0.0256x^{16} + 0.0911x^{19}
+0.04x^{39} + 0.0117x^{51}
+0.0189x^{57} + 0.0112x^{62} + 0.0684x^{76}$

$\rho (x) = 0.2{x^3}+0.8{x^4}$

------------------------------------------------------------------------------------------------------------------

Code 5:  $1-R_2=0.818$

$\lambda (x) = 0.3038x + 0.1731x^2 + 0.0671x^4
+0.0123x^5 + 0.1341x^6 + 0.0314x^{12}
+0.0108x^{14} + $

$ 0.0256x^{16} + 0.0911x^{19}
+0.04x^{39} + 0.0117x^{51}
+0.0189x^{57} + 0.0112x^{62} + 0.0684x^{76}$

$\rho (x) = 0.2{x^3}+0.8{x^4}$

------------------------------------------------------------------------------------------------------------------

Code 11:   $1-R_2=0.294$

$\lambda (x) = 0.1392x + 0.2007{x}^2 + 0.2522{x}^6
+0.0134{x}^{11} + 0.171{x}^{17} + 0.0424{x}^{31}
+0.0855{x}^{41} +$

$ 0.0953{x}^{49}$

$\rho (x) = 0.3{x^{16}}+0.7{x^{17}}$

------------------------------------------------------------------------------------------------------------------

Code 13:   $1-R_2=0.92$

$\lambda (x) = 0.4051x + 0.1716x^2 + 0.0995x^4
+0.0447x^5 + 0.0379x^6 + 0.0612x^{10}
+0.0189x^{14} +$

$ 0.0333x^{16} + 0.0026x^{17}
+0.0128x^{20} + 0.0435x^{28} + 0.0075x^{50}
+0.0123x^{52} + 0.0258x^{62} + $

$0.0065x^{63} +0.0166x^{71}$

$\rho (x) = 0.4{x^{2}}+0.6{x^{3}}$

------------------------------------------------------------------------------------------------------------------

Code 14:   $1-R_2=0.605$

$\lambda (x) = 0.2366x + 0.3138x^3
+0.0715x^4 +0.1707x^{10} + 0.0005x^{11}  + 0.0002x^{18}
+0.0002x^{19} +$

$ 0.0002x^{20} + 0.0003x^{21}
+0.0007x^{22} + 0.0183x^{23} + 0.1854x^{24}
+0.0016x^{25} $

$\rho (x) = 0.9{x^{6}}+0.1{x^{7}}$

------------------------------------------------------------------------------------------------------------------

For the parity-check matrices of the LDPC codes in Tables \ref{t2reg} and \ref{t4reg} we have used $(9,10)$, $(7,8)$, and $(3,10)$-regular codes.
Note that, in order to get the exact rate of $1-R_2$, some rows of the parity-check matrices are randomly removed.

\end{document}